\documentclass[superscriptaddress,prd,aps,preprint]{revtex4}

\usepackage{graphicx,amsfonts,amsmath,color,comment}
\renewcommand{\a}{\alpha}

\renewcommand{\b}{\beta}
\newcommand{\q}{\theta}

\newcommand{\pa}{\partial}
\newcommand{\hf}{\frac{1}{2}}

\begin{document}

\title{Perturbative aspects of the supersymmetric three-dimensional massive QED}

\author{A. C. Lehum}
\email{lehum@ufpa.br}
\affiliation{Faculdade de F\'{i}sica, Universidade Federal do Par\'{a}\\ 66075-110, Bel\'{e}m, PA, Brazil}

\author{J. R. Nascimento}
\email{jroberto@fisica.ufpb.br}
\affiliation{Departamento de F\'{\i}sica, Universidade Federal da Para\'{\i}ba\\
 Caixa Postal 5008, 58051-970, Jo\~ao Pessoa, Para\'{\i}ba, Brazil}

\author{A. C. Pina Neto}
\email{aldo.neto@icen.ufpa.br}
\affiliation{Faculdade de F\'{i}sica, Universidade Federal do Par\'{a}\\ 66075-110, Bel\'{e}m, PA, Brazil}

\author{A. Yu. Petrov}
\email{petrov@fisica.ufpb.br}
\affiliation{Departamento de F\'{\i}sica, Universidade Federal da Para\'{\i}ba\\
 Caixa Postal 5008, 58051-970, Jo\~ao Pessoa, Para\'{\i}ba, Brazil}

%\begin{abstract} We formulate massive supersymmetric QED and calculate lower quantum corrections within it. \end{abstract}

\begin{abstract}
We perform the study of perturbative aspects of a three-dimensional supersymmetric Maxwell-Chern-Simons-Proca theory minimally coupled to scalar superfields. Using the superfield formalism, we derive the propagators for both gauge and matter superfields and compute the leading quantum corrections to the effective action. The presence of the Proca-like term explicitly breaks gauge invariance, modifying the structure of the gauge superfield propagator and leading to an essentially new form of quantum contributions in comparison with the usual QED. We analyze the Feynman diagrams that contribute to the quadratic part of the effective action, obtaining corrections to both the kinetic and mass terms of the scalar superfields. %By expressing the results in terms of integral representations, we examine the dependence of the effective action on the Proca parameter \( \rho \). 
Furthermore, we discuss the UV behavior of the model, considering its renormalization properties and the possibility of perturbative finiteness to all loop orders, similar to supersymmetric QED$_3$. Finally, we highlight potential applications of this model in condensed matter systems and possible connections with modified supersymmetric electrodynamics and dualities in lower-dimensional theories.
\end{abstract}

\maketitle

\section{Introduction}

Three-dimensional field theories acquired a notable interest within modern theoretical physics, not only due to their relative simplicity, but also due to their application to possible analogue models of condensed matter, especially graphene \cite{Geim}.  In this context, it is certainly important to study supersymmetric extensions of such models as they could display better renormalization properties; in particular, a possible supersymmetric model for graphene has been formulated in \cite{Abreu}, with further developments presented in \cite{Oiko,Gupta,ourpseudo}. A natural direction in this area involves studying various three-dimensional supersymmetric field models, particularly within the superfield approach, which is recognized as one of the most effective tools for perturbative calculations (for a general description of the three-dimensional superfield formalism see, e.g. \cite{SGRS,ourbook}). 

In this framework, supersymmetric massive QED plays a significant role, as it provides a setting to explore scenarios in which gauge symmetry is broken. In ~\cite{Lehum:2009vui}, a scenario was presented in which radiative corrections can generate a mass for the gauge superfield, leading to spontaneous breaking of the gauge symmetry. It was shown that a Proca-like term naturally emerges as a radiative correction. Consequently, introducing a Proca-like term from the outset constitutes a well-motivated and consistent approach in the study of supersymmetric gauge theories. Furthermore, mass terms can also be generated radiatively in certain phases of supersymmetric CP$^{(N-1)}$ models~\cite{cpn,Ferrari:2006xx}, reinforcing their relevance in different supersymmetric frameworks.

Thus, in this work, we intend to study the perturbative aspects of a three-dimensional supersymmetric Maxwell-Chern-Simons-Proca theory minimally coupled to scalar superfields. Using the superfield formalism, we derive the propagators for both gauge and matter superfields and compute the leading quantum contributions to the effective action. 

The paper is organized as follows. In Sec.~\ref{sec:model}, we define the model and derive the corresponding propagators for the gauge and scalar superfields. In Sec.~\ref{sec:quantum_corrections}, we compute the quantum corrections to the effective action, focusing on the modifications to the kinetic and mass terms induced by radiative effects. In Sec.~\ref{sec:discussion}, we discuss the implications of our results and compare them with previous studies, particularly with respect to the role of the Proca-like term, and finally present our final remarks and suggest possible future directions for research.

Throughout this paper, we employ the natural unit setting $c = \hbar = 1$ and adopt the space-time signature $(-,+,+)$. Additionally, we follow the superspace conventions established in Refs.~\cite{SGRS,ourbook}.

\section {Definition of the model and the propagators}\label{sec:model}

As it is known (see e.g. \cite{SGRS}), the supersymmetric $3D$ Maxwell-Chern-Simons theory is described by the action ($\alpha=1,2$):
\begin{eqnarray}
	\label{action0}
	S&=&\int d^5 z \left[\frac{1}{2} W^\alpha W_\alpha +\frac{M}{2}W^{\alpha} A_{\alpha}+{\cal L}_m\, \right], 
	%\label{2n}
\end{eqnarray}
where
$W_\beta =\frac{1}{2}D^\alpha D_\beta A_\alpha$ 
is the standard spinor superfield strength \cite{SGRS} constructed on the base of the gauge superfield $A_\alpha$ and $d^5z=d^3x d^2\theta$. The theory (\ref{action0}) is invariant under  usual supergauge transformations $\delta A_{\alpha}=D_{\alpha}\xi$,  where $\xi$ is the scalar superfield parameter. The ${\cal L}_m$ is the relevant matter Lagrangian.

It is instructive to present the component form of the action given in Eq.~(\ref{action0}):
\begin{eqnarray}
S=\int d^3 x \left[-(\frac{1}{2}f^{\alpha\beta}f_{\alpha\beta}+\lambda^{\alpha}i\partial_{\alpha\beta}\lambda^{\beta})+M (V^{\alpha\beta}f_{\alpha\beta}-\lambda^{\alpha}\lambda_{\alpha})\right]+{\cal L}_m,
\end{eqnarray}
where $V_{\alpha\beta}=\gamma^m_{\alpha\beta}A_m$ denotes the bispinor representation of the vector field $A_m$, $ f_{\alpha\beta} = \gamma^m_{\alpha\beta} \epsilon_{mln} F^{ln} $ is the bispinor representation of the dual field strength tensor $F_{mn}$, and $\lambda_{\beta}$ is the spinor superpartner of the gauge field. The component fields of the superfield $A_{\alpha}$ are defined as
\begin{eqnarray}
\label{spi}
\chi_{\alpha}(x)&=&A_{\alpha}(x,\theta)|;\quad
B(x)=\frac{1}{2}D^{\alpha}A_{\alpha}(x,\theta)|;\quad
V_{\alpha\beta}(x)=-\frac{i}{2}D_{(\alpha}A_{\beta)}(x,\theta)|;\nonumber\\
\lambda_{\alpha}&=&\frac{1}{2}D^{\beta}D_{\alpha}A_{\beta}(x,\theta)|.
\end{eqnarray}
The next step consists in breaking the gauge symmetry by introducing a Proca-like additive term, defined as ${\cal L}_P = \frac{\rho}{2} M^2 A^{\alpha} A_{\alpha}$, where $ \rho$ is a dimensionless parameter. The resulting free action then reads
\begin{eqnarray}
	\label{action1}
	S &=& \int d^5 z \left[\frac{1}{2} W^\alpha W_\alpha + \frac{M}{2} W^{\alpha} A_{\alpha} + \frac{\rho}{2} M^2 A^{\alpha} A_{\alpha} \right].
\end{eqnarray}
The corresponding component form of the action is given by
\begin{eqnarray}
	S &=& \int d^3 x \left[-\left(\frac{1}{2}f^{\alpha\beta}f_{\alpha\beta} + \lambda^{\alpha} i \partial_{\alpha\beta} \lambda^{\beta} \right) 
	+ M \left( V^{\alpha\beta} f_{\alpha\beta} - \lambda^{\alpha} \lambda_{\alpha} \right) \right. \nonumber \\
	&& \left. + \frac{\rho M^2}{2} \left( V^{\alpha\beta} V_{\alpha\beta} - B^2 + 2 (i \partial^{\alpha\beta} \chi_{\beta} - \lambda^{\alpha}) \chi_{\alpha} \right) \right] + {\cal L}_m.
\end{eqnarray}

\begin{comment}
The corresponding propagator of the spinor superfield $A_{\alpha}$ is
\begin{eqnarray}
	<A_{\alpha}(-k,\theta_1)A_{\beta}(k,\theta_2)>&=&-\frac{2i}{k^2}\left[
\frac{D^2D_{\alpha}D_{\beta}}{\rho k^2}-\frac{D^2D_{\beta}D_{\alpha}}{-k^2+MD^2+\rho M^2}
	\right]\delta_{12}\nonumber\\
    &=& -\frac{2i}{k^2} \left[\frac{D^{2} D^{\alpha} D^{\beta}}{\rho k^2} + \frac{(k^2- \rho M^2 +M D^2) D^2 D^{\beta} D^{\alpha}}{M^2 k^2 + (\rho M^2 - k^2)^2}   \right] \delta_{1 2},
	\end{eqnarray}
\noindent where $\delta_{1 2}=\delta(\theta_1-\theta_2)$. 
We note that this propagator in the UV limits behaves as $\frac{1}{k^2}$, just as in the usual three-dimensional super-QED. Hence, the renormalization behavior of the theory is the same as in the massless case. Also, we note that this propagator yields real poles only for $\rho\leq 1/4$, otherwise the poles will be complex, making the values of $\rho>1/4$ physically inconsistent. 
Besides this, we note that this propagator is not well-defined at $\rho=0$. Such a singularity is a natural analogue of the well-known  fact that the massless limit in four-dimensional Proca theory is ill-defined, which is reasonable since massive and massless vector fields possess different numbers of degrees of freedom, and the same situation occurs in our case.
However, it should be noted that despite this, our quantum corrections are well-defined in the limit $\rho \to 0$.
\end{comment}

The propagator for the spinor superfield $ A_{\alpha}$ takes the form
\begin{eqnarray}
	\langle A_{\alpha}(-k,\theta_1) A_{\beta}(k,\theta_2) \rangle &=& -\frac{2i}{k^2} \left[
	\frac{D^2 D_{\alpha} D_{\beta}}{\rho k^2} - \frac{D^2 D_{\beta} D_{\alpha}}{-k^2 + M D^2 + \rho M^2}
	\right] \delta_{12} \nonumber \\
	&=& -\frac{2i}{k^2} \left[ \frac{D^2 D_{\alpha} D_{\beta}}{\rho k^2} + \frac{(k^2 - \rho M^2 + M D^2) D^2 D_{\beta} D_{\alpha}}{M^2 k^2 + (\rho M^2 - k^2)^2} \right] \delta_{12},
\end{eqnarray}
\noindent where \( \delta_{12} = \delta(\theta_1 - \theta_2) \).

It is worth noting that, in the UV limit, this propagator behaves as $ 1/k^2$, similar to the behavior in conventional three-dimensional massless super-QED. Therefore, the renormalization properties of the theory remain unchanged relative to the massless case. Furthermore, the propagator exhibits real poles only for \( \rho \leq 1/4 \); for \( \rho > 1/4 \), the poles become complex, rendering such values physically inconsistent.

%\noindent
It should also be emphasized that the propagator becomes ill-defined in the limit $\rho \to 0$. This singular behavior is analogous to the well-known discontinuity in the massless limit of the four-dimensional Proca theory, which arises due to the mismatch in the number of degrees of freedom between massive and massless vector fields. A similar situation occurs in the present model. Nevertheless, we stress that despite the singularity in the propagator, all quantum corrections remain well-defined in the limit $\rho \to 0$.

Afterwards, we couple our theory to the scalar matter. Since our model is not gauge invariant anymore, new interactions can be introduced. However, as a first step, we consider the simple case, that is, the usual coupling with $N$ scalar superfields through the same action employed in various papers on scalar super-QED, for example, \cite{cpn}:
\begin{eqnarray}
\label{acmat}
S_m &=& \int d^{5}z  \left[\frac12(\overline{\nabla^{\alpha} \phi_{a}})\nabla_{\alpha} \phi_{a}+m\bar{\phi}_a\phi_a\right]\nonumber\\
&=&\int d^5 z
	\Big[-\bar{\phi}_a(D^2-m)\phi_a+i\frac{g}{2} (\bar{\phi}_aA^\alpha D_\alpha \phi_a-
	D_\alpha \bar{\phi}_aA^\alpha \phi_a)+
	\frac{g^2}{2} \bar{\phi}_aA^\alpha A_\alpha\phi_a\Big],
\end{eqnarray}
\noindent where a sum over the repeated isotopic indices $a=1,\cdots,N$ is understood. The gauge supercovariant derivative $\nabla_{\alpha} = D_{\alpha}+i g A_{\alpha}$ is introduced to ensure a minimal coupling between the matter superfield and the spinor one. Although the model is no longer gauge invariant due to the presence of the Proca-like term, this breaking is soft, meaning that the gauge symmetry is only explicitly broken at the mass scale introduced by $\rho M^2$. Consequently, the introduction of minimal coupling remains a well-motivated approach, as it preserves supersymmetric interactions while incorporating gauge interactions in a controlled manner.

The component form of this action is
\begin{eqnarray}
S_m&=&\int d^3 x
\big[F_a\bar{F}_a-i\bar{\psi}_{a\a}\pa^{\a\b}\psi_{a\b}+
\bar{\varphi}_a\Box \varphi_a-m(2\bar{\psi}^{\a}_a\psi_{a\a}+\phi_a\bar{F}_a+\bar{\phi}_aF_a)\nonumber\\&+&
i V^{\a\b}\varphi_a\stackrel{\leftrightarrow}{\pa}_{\a\b}\bar{\varphi}_a
-\frac{1}{2}\varphi_a V^{\a\b}V_{\a\b}\bar{\varphi}_a-
V^{\a\b}(\psi_{a\a}\bar{\psi}_{a\b}+\bar{\psi}_{a\a}\psi_{a\b})
+\lambda^{\a}(\varphi_a\bar{\psi}_{a\a}+\bar{\varphi}_a\psi_{a\a})
-\nonumber\\&-&
(\psi^{\a}_a\bar{F}_a-\bar{\psi}^{\a}_a F_a)\chi_{\a}-
(\pa_{\a\b}\psi^{\a}_a\bar{\varphi}_a-\pa_{\a\b}\bar{\psi}^{\a}_a\varphi_a+
\pa_{\a\b}\bar{\varphi}_a\psi^{\a}_a+\pa_{\a\b}\varphi_a\bar{\psi}^{\a}_a)\chi^{\b}-
\nonumber\\&-&\frac{1}{2}(\varphi_a B^2\bar{\varphi}_a-
\varphi_a\bar{\varphi}_a\chi^{\a}\lambda_{\a}+\chi^{\a}\chi_{\a}
(F_a\bar{\varphi}_a+\varphi_a\bar{F}_a))\big].
\end{eqnarray}
The free propagator of the scalar superfields is given by
\begin{eqnarray}
	<\bar{\phi}_a(-k,\theta_1)\phi_b(k,\theta_2)>=i\delta_{ab}\frac{D^2+m}{k^2+m^2}
	\delta_{12}\,.
\end{eqnarray}

Another possible approach involves the introduction of non-minimal couplings, for example, $f(\phi_a\bar{\phi}_a)A^{\alpha}A_{\alpha}$ and similar vertices, allowed now due to the absence of the gauge symmetry. However, for the first step, we restrict our analysis to the model with only minimal coupling (\ref{acmat}),  which, as can be seen, is better from a renormalization point of view.

\section{Radiative corrections}\label{sec:quantum_corrections}

Quantum corrections play a crucial role in understanding the behavior of supersymmetric field theories, particularly in the presence of gauge-breaking terms. In this section, we start with discussing the matter loop corrections to the spinor superfield effective action and compute the leading radiative corrections to the effective action of the matter superfields in the three-dimensional supersymmetric Maxwell-Chern-Simons-Proca model. The presence of the Proca-like term modifies the structure of the gauge superfield propagator, leading to quantum contributions essentially different from those ones arising in the usual gauge Maxwell-Chern-Simons theory (see, e.g. \cite{Ferrari:2006xx,Ferrari:2007mh,Ferrari:2007vv}).  We discuss the two-point functions of spinor superfields and of scalar (matter) superfields.  Whereas in the first case, the standard spinor-scalar couplings yield the same contributions as for the presence of the gauge symmetry \cite{Ferrari:2006xx,Ferrari:2007mh,Ferrari:2007vv}, in the matter sector, we will arrive at essentially new results. Explicitly, in the last case we analyze the two main Feynman diagrams that contribute to the quadratic part of the effective action of the matter: one associated with the correction to the mass term and another that influences the kinetic and mass terms. Let us start with the $A_{\alpha}$-dependent contributions to the effective action.

\begin{figure}[htbp] 
	\begin{center} 
		\includegraphics[width={10cm}]{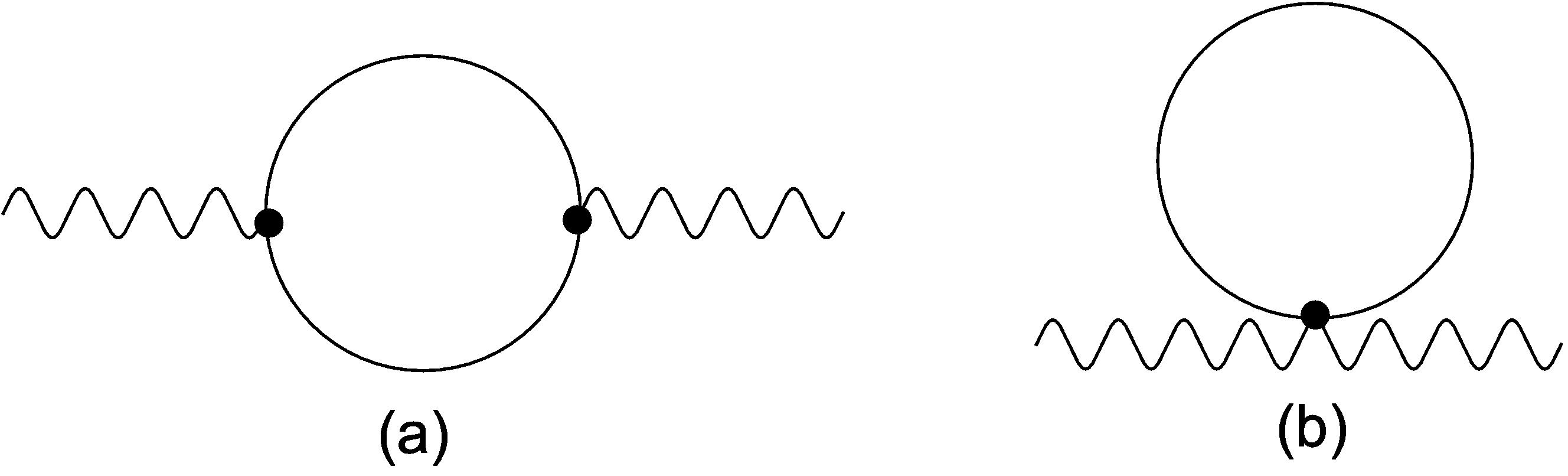}  
	\end{center} 
	\caption{Contributions to the two-point function of the gauge superfield.}
	\label{Fig:diagrams1} 
\end{figure}

It is clear that in the case of the usual coupling given by (\ref{acmat}) and external vector lines (see Fig. \ref{Fig:diagrams1}) we reproduce the well-known result for the simple nonlocal generalization of the Maxwell-Chern-Simons (MCS) action (cf. \cite{ourpseudo,cpn,ourbook}).
Explicitly, for $N$ matter superfields, the contribution of the supergraph given by Fig. 1a is (cf. \cite{ourbook}):
\begin{eqnarray}
\label{s10}
iS_{1a}(p)&=&-\frac{N}{2}\int d^2\q \int \frac{d^3k}{(2\pi)^3} 
I(k,p)
\nonumber\\&\times&\Big[
(k^2+m^2)C_{\a\b} A^{\a}(-p,\q)A^{\b}(p,\q)
+(k_{\a\b}+mC_{\a\b})(D^2A^{\a}(-p,\q)) A^{\b}(p,\q)
\nonumber\\&+&\hf D^{\gamma}D^{\a}A_{\a}
(-p,\theta)(k_{\gamma\b}+mC_{\gamma\b})A^{\b}(p,\q)
\Big].
\end{eqnarray}
The contribution of the supergraph gven by Fig. 1b looks like:
 \begin{eqnarray}
\label{s20}
iS_{1b}(p)&=&\frac{N}{2}\int\frac{d^3k}{(2\pi)^3}\frac{1}{(k+p)^2+m^2}
C_{\a\b} A^{\a}(-p,\q)A^{\b}(p,\q).
\end{eqnarray}
%Both these results differ from zero being superficially divergent. Actually, they are finite due to the effects of dimensional regularization. Summing these results and performing the integration, we arrive at \cite{ourbook}:

\noindent
Both contributions are non-vanishing and superficially divergent. However, they turn out to be finite as a consequence of dimensional regularization. By summing the two expressions and carrying out the integration, we obtain \cite{ourbook}: 
\begin{eqnarray}
	\Sigma&=&\frac{Ng^2}{4}\int d^2\theta\frac{d^3p}{(2\pi)^3} \frac{1}{4\pi\sqrt{p^2}}\arctan \left(\frac{1}{2}\sqrt{p^2/m^2}\right) \times\nonumber\\ &\times&
	\left[W^{\alpha}(-p)W_{\alpha}(p)+2mW^{\alpha}(-p)A_{\alpha}(p)\right],\label{action1a}
\end{eqnarray}
so that in the IR limit we recover the simple MCS action. Actually, this is reasonable since we have a gauge invariant matter sector.

Furthermore, we compute the quadratic part of the effective action for the matter superfield. In this case, the diagram in Fig.~\ref{Fig:diagrams2} (b) corresponds to the contribution arising solely from the second term in the gauge superfield propagator, leading to a radiative correction to the mass term:
\begin{eqnarray}\label{eq_fig2.2}
	\tilde\Gamma_b &=& -2 g^2 M \phi_a\bar{\phi}_a\int\frac{d^3k}{(2\pi)^3}\frac{1}{(-k^2+\rho M^2)^2+M^2k^2}= -2 g^2 M I_1 \phi_a\bar{\phi}_a,
\end{eqnarray}
where the integral \( I_1 \) is evaluated in the Appendix. It is important to note that \( \tilde\Gamma_b \) is related to the effective action via the relation
\begin{equation}
    \Gamma_b = \int d^2\theta\frac{d^3p}{(2\pi)^3}\tilde{\Gamma}_b.
\end{equation}

At the same time, the diagram in Fig.~\ref{Fig:diagrams2} (a) is more intricate and contributes both to the mass term and to the kinetic term. It is given by
\begin{eqnarray}
	\Gamma_a&=&\frac{g^2}{2}\int d^{5}z_1\,d^{5}z_2\,<A^{\alpha}(1)A^{\beta}(2)>\times\nonumber\\&\times&
	\Big[
	D_{\alpha}\phi(1)\bar{\phi}(2)<\bar{\phi}(1)D_{\beta}\phi(2)>+<D_{\alpha}\phi(1)\bar{\phi}(2)>\bar{\phi}(1)D_{\beta}\phi(2)-\nonumber\\&-&<D_{\alpha}\bar{\phi}(1)D_{\beta}\phi(2)>\phi(1)\bar{\phi}(2)-D_{\alpha}\bar{\phi}(1)D_{\beta}\phi(2)<\phi(1)\bar{\phi}(2)>-\nonumber\\
	&-&\bar{\phi}(1)\phi(2)<D_{\alpha}\phi(1)D_{\beta}\bar{\phi}(2)>-<\bar{\phi}(1)\phi(2)>D_{\alpha}\phi(1)D_{\beta}\bar{\phi}(2)+\nonumber\\&+&
	<D_{\alpha}\bar{\phi}(1)\phi(2)>\phi(1)D_{\beta}\bar{\phi}(2)+D_{\alpha}\bar{\phi}(1)\phi(2)<\phi(1)D_{\beta}\bar{\phi}(2)>
	\Big],
\end{eqnarray}
\noindent where $\phi(n)$ ($A^\alpha(n)$) is a shorthand notation for $\phi(z_n)$ ($A^\alpha(z_n)$).

By substituting the expressions for the propagators and performing the corresponding \( D \)-algebra computations with the aid of the Mathematica package \texttt{SusyMath}~\cite{Ferrari:2007sc}, we obtain
\begin{eqnarray}
   \tilde\Gamma_{a} &=& g^2 \int \frac{d^3k}{(2\pi)^3} \frac{1}{(k+p)^2 + m^2} \Bigg[ \frac{-4 p^2 \bar{\phi}_a D^2 \phi_a + 4 k^2 \bar{\phi}_a (D^2-2m) \phi_a}{\rho (k^2)^2} + \nonumber \\ 
   && +\frac{ 12 k^2 \bar{\phi}_a (D^2-M/3) \phi_a 
   + M ~\bar{\phi}_a [p^2+12(m-\rho M)D^2]\phi_a}{M^2 k^2 + (\rho M^2 - k^2)^2} \Bigg]\nonumber\\
   &=&g^2\int \frac{d^3k}{(2\pi)^3} \frac{1}{k^2 + m^2} \Bigg[ \frac{4 \bar{\phi}_a (D^2-2m) \phi_a}{\rho~k^2} + \frac{12 k^2 \bar{\phi}_a (D^2-M/3) \phi_a}{M^2 k^2 + (\rho M^2 - k^2)^2} \nonumber \\ 
   &&+ \frac{12 M (m-\rho M)~\bar{\phi}_a D^2\phi_a}{M^2 k^2 + (\rho M^2 - k^2)^2} \Bigg]+\mathcal{O}(p^2),
\end{eqnarray}
where, in the final step, in order to consider the low-energy leading contribution, we have expanded \(\tilde\Gamma_{a}\) around $p^2 \approx 0$.

Expressing $\tilde\Gamma_a$ in terms of the integrals defined in the Appendix, we obtain
\begin{eqnarray}
   \tilde\Gamma_{a}
   &=& \frac{4 g^2 I_4}{\rho}~ \bar{\phi}_a (D^2-2m) \phi_a
   + 12 g^2 I_3~\bar{\phi}_a (D^2-M/3) \phi_a \nonumber \\ 
   &&
   + 12 g^2 M (m-\rho M) I_2~\bar{\phi}_a D^2\phi_a
   +\mathcal{O}(p^2).
\end{eqnarray}

By adding both contributions given by Fig.\ref{Fig:diagrams2}, the quadratic part of the effective action for the scalar superfield, in the regime \(0<\rho\leq 1/4\), is given by
\begin{eqnarray}\label{scalar_se_01}
   \tilde\Gamma &=&\tilde\Gamma_a+\tilde\Gamma_b \nonumber\\
   &=&\frac{ig^2}{m\pi\rho}\bar\phi_a (D^2-2m) \phi_a 
   + \frac{6ig^2}{(m+M)\pi}\bar\phi_a \left[D^2-\frac{(m+3M)}{12} \right] \phi_a\nonumber\\
   &&
   -\frac{6ig^2\rho M^2}{m(m+M)^2\pi} \bar\phi_a (D^2+m/6) \phi_a +\mathcal{O}(\rho^2), ~~~~\mathrm{for}~0<\rho\leq 1/4.
\end{eqnarray}
\noindent Similarly, in the regime \(\rho<0\), the effective action takes the form
\begin{eqnarray}\label{scalar_se_02}
   \tilde\Gamma 
   &=&\frac{ig^2}{m\pi\rho}\bar\phi_a (D^2-2m) \phi_a 
   + \frac{6ig^2}{(m+M)\pi}\bar\phi_a \left[D^2-\frac{(m+3M)}{12} \right] \phi_a\nonumber\\
   &&
   +\frac{6ig^2\rho (m+2M)}{(m+M)^2\pi} \bar\phi_a \left[D^2-\frac{(m+M)^2}{6(m+2M)} \right] \phi_a +\mathcal{O}(\rho^2), ~~~~\mathrm{for}~\rho<0.
\end{eqnarray}

Combining the contributions from both diagrams in Fig.~\ref{Fig:diagrams2}, we have derived the quadratic part of the effective action for the scalar superfield in different regimes of \( \rho \), Eqs. \eqref{scalar_se_01} and \eqref{scalar_se_02}. In the range $0 < \rho \leq 1/4$, the effective action acquires corrections that modify both the kinetic and the mass terms, as shown in \eqref{scalar_se_01}. Similarly, for $\rho < 0$, a distinct structure emerges due to the  change of sign in the relevant terms (we note again that the case $\rho>1/4$ is physically inconsistent). These results highlight the dependence of the effective action on the Proca-like parameter $\rho$, which controls the modifications induced by radiative corrections. 

\section{Final Remarks}\label{sec:discussion}

In this work, we have formulated a three-dimensional supersymmetric massive QED by incorporating a Proca-like term into the gauge sector of the theory. Using the superfield formalism, we derived the propagators for the gauge and matter superfields and obtained the one-loop contributions to the effective action.

We calculated the leading-order radiative corrections to the free action of the scalar superfield, considering different regimes of the Proca-like parameter $\rho$. Our results reveal that, for $0 < \rho \leq 1/4$, the effective action receives modifications both to the kinetic and mass terms, with corrections proportional to $\rho$. On the other hand, for $\rho < 0$, a distinct structure emerges due to sign changes in key contributions, modifying the behavior of the effective action. These findings highlight the role of the parameter $\rho$ in controlling the influence of quantum corrections in the model.

In a previous paper by some of us \cite{Lehum:2009vui} it has been demonstrated how radiative corrections can induce a mass for the gauge superfield, leading to spontaneous gauge symmetry breaking. This work has inspired us to consider a massive scalar superfield minimally interacting with a gauge superfield governed by the Maxwell-Chern-Simons-Proca action in three dimensions. However, while in our case the gauge field mass is introduced explicitly via a Proca-like term, in Ref. \cite{Lehum:2009vui}, it arises dynamically. Both approaches explore mechanisms of mass generation for gauge fields in supersymmetric three-dimensional theories, albeit through distinct frameworks. Additionally, the inclusion of mass terms for gauge fields in three dimensions is associated with parity anomaly effects and dualities in supersymmetric gauge theories (see, e.g. \cite{Ferrari:2006vy} and references therein). The introduction of Chern-Simons or Proca terms can influence duality properties and modify the structure of three-dimensional supersymmetric gauge theories. These connections are relevant for understanding non-perturbative effects and the dynamics of lower-dimensional gauge theories.

The approach presented in this work provides a framework for studying supersymmetric gauge theories with explicit gauge symmetry breaking. Future investigations may focus on exploring the renormalization properties of the model and analyzing the impact of higher-order quantum corrections. Regarding the UV behavior of the model, it would be interesting to verify whether it remains perturbatively finite for all loop orders, similar to SUSY QED$_3$~\cite{Ferrari:2007mh}. Additionally, it would be valuable to identify potential applications in condensed matter systems, where three-dimensional gauge theories play a crucial role in effective descriptions of planar physics (see e.g. \cite{Abreu,Oiko,Gupta} and references therein). Furthermore, the presence of the Proca term suggests potential connections with modified supersymmetric electrodynamics and dualities in lower-dimensional theories~\cite{Gomes:2004ka}. In addition, since the Maxwell-Chern-Simons-Proca theory is not gauge invariant, it is possible to introduce a more generic class of scalar-vector couplings and study their perturbative impacts.

To close this work, we note that our results provide a foundation for further studies on supersymmetric field theories with massive gauge fields and their implications for both theoretical and phenomenological applications. We expect to consider these applications, as well as more generic scalar-vector couplings, in our forthcoming papers.

\vspace{.5cm}
{\bf Acknowledgments.}
A. C. P. N. is partially
supported by Coordena\c{c}\~ao de Aperfei\c{c}oamento de Pessoal de N\'ivel Superior (CAPES). The work of A. C. L. has been partially supported by the CNPq project No. 404310/2023-0. A. Yu. P. has been partially supported by the CNPq project No. 303777/2023-0. 

\appendix
\section*{Appendix: Evaluation of integrals}

\begin{figure}[htbp] 
	\begin{center} 
		\includegraphics[width={10cm}]{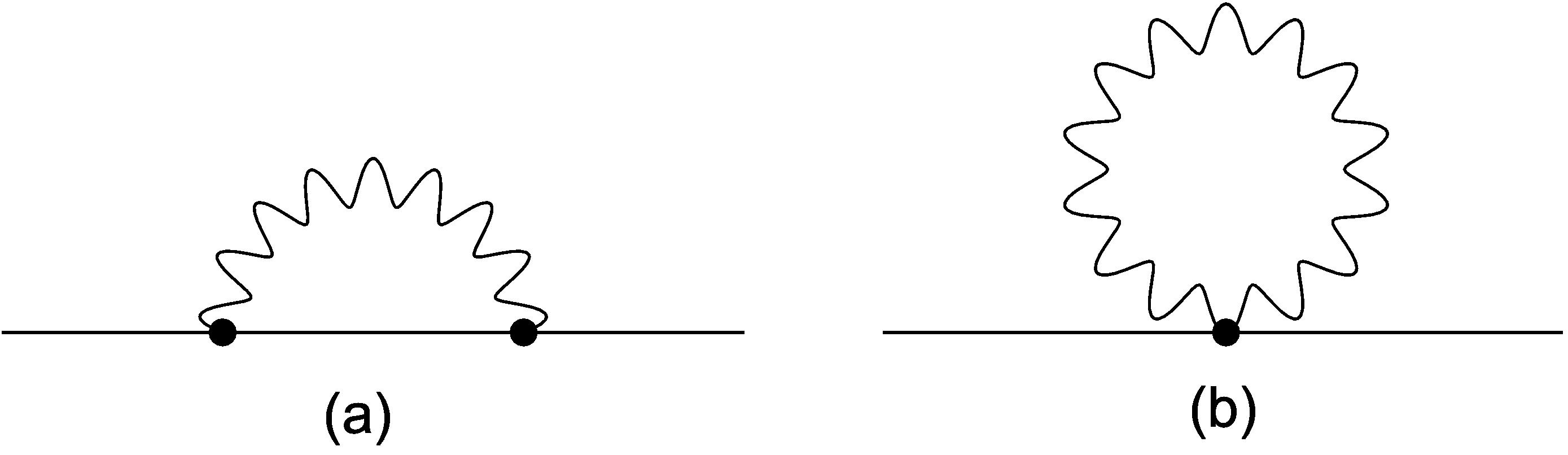}  
	\end{center} 
	\caption{Contributions to the two-point function of the matter superfield.}
	\label{Fig:diagrams2} 
\end{figure}

The first integral we need to compute is the one associated with the self-energy diagram of the scalar superfield, depicted in Figure \ref{Fig:diagrams2}.2, and appearing in Eq.~\eqref{eq_fig2.2}:
\begin{eqnarray}
	I_1 &=& \int\frac{d^3k}{(2\pi)^3}\frac{1}{(-k^2+\rho M^2)^2+M^2k^2}.
\end{eqnarray}
Performing the Wick rotation, the integral $I_1$ can be rewritten as
\begin{eqnarray}
	I_1 &=& \int_0^{\infty}\frac{4\pi i k_E^2 dk_E}{(2\pi)^3}\frac{1}{(-k_E^2+\rho M^2)^2+M^2k_E^2},\nonumber\\
    &=& \left\{\begin{array}{cc}
          \frac{\sqrt{2} i \left( \sqrt{\frac{2\rho - \sqrt{1 - 4\rho} - 1}{4\rho - 1}} (4\rho - 1) + \sqrt{(4\rho - 1)(2\rho + \sqrt{1 - 4\rho} - 1)} \right)}{8\pi M (4\rho - 1)}=\dfrac{i}{4\pi M}+\mathcal{O}(\rho^2),
& \text{for } 1/4\ge\rho>0, \\[10pt]
 \frac{\sqrt{2} i \left( \sqrt{\frac{2\rho - \sqrt{1 - 4\rho} - 1}{4\rho - 1}} - \sqrt{\frac{2\rho + \sqrt{1 - 4\rho} - 1}{4\rho - 1}} \right)}{8\pi M}
=\dfrac{i}{4\pi M}(1+2\rho)+\mathcal{O}(\rho^2),
        & \text{for } \rho<0,
    \end{array} \right.
\end{eqnarray}
where $k_E$ is the modulus of the Euclidean three-momentum. In this work, we consider $m>0$  and $M>0$, while $\rho$ can take positive or negative values. 
%{\bf It is easy to check that the case $\rho>1/4$ implies complex poles of the propagator and hence is physically inconsistent.}

The integral $I_2$ can be cast as
\begin{eqnarray}
	I_2 &=& \int \frac{d^3k}{(2\pi)^3} \frac{1}{(k^2 + m^2)}\frac{1}{M^2 k^2 + (\rho M^2 - k^2)^2}\nonumber\\
    &=& \left\{\begin{array}{cc}
              \dfrac{i}{4 \pi m M (m+ M)} -\dfrac{i~M \rho}{4\pi m^2 (m + M)^2}
+\mathcal{O}(\rho^2),
& \mathrm{for~}1/4\ge\rho>0 \\
 \dfrac{i}{4 \pi m M (m+M)} + \dfrac{i~(2 m^2 + 4 m M + M^2) \rho}{4\pi m^2 M (m + M)^2}
+\mathcal{O}(\rho^2),
        & \mathrm{for~}\rho<0.
    \end{array} \right.
\end{eqnarray}

The integral $I_3$ is given by
\begin{eqnarray}
	I_3 &=& \int \frac{d^3k}{(2\pi)^3} \frac{k^2}{(k^2 + m^2)}\frac{1}{M^2 k^2 + (\rho M^2 - k^2)^2}\nonumber\\
    &=& \frac{i \left( m + M(1+\rho)\right)}{4 \pi (M + m)^2}
+\mathcal{O}(\rho^2),~\mathrm{for~}\rho\le 1/4.
\end{eqnarray}

Finally, the integral $I_4$ is
\begin{eqnarray}
	I_4 &=& \int \frac{d^3k}{(2\pi)^3} \frac{1}{k^2(k^2 + m^2)}=\frac{i}{4 \pi m}.
\end{eqnarray}

\vspace*{2mm}

\vspace*{2mm}

\vspace*{2mm}

\end{document}